\documentstyle[aps,prl,twocolumn]{revtex}
\begin{document}

\draft \wideabs %
{
\title{Soliton--Magnon Scattering in Two--Dimensional Isotropic Ferromagnets}
\author{B.~A.~Ivanov$^*$$^\dag$, V.~M.~Muravyov$^*$$\ddag$, and
D.~D.~Sheka$^\dag$\cite{Sheka:e-mail}}
\address{$^*$ Institute of Magnetism, National Academy of Sciences of 
the Ukraine, Kiev 252142, Ukraine}
\address{$^\dag$ Radiophysics Faculty, Taras Shevchenko Kiev University,
Kiev 252127, Ukraine}
\address{$^\ddag$ Faculty of Physics and Mathematics, The National 
Pedagogical University by Dragomanov, Kiev 252030, Ukraine}
\date{May 21, 1999}
\maketitle

\begin{abstract}
It is studied the scattering of magnons by the 2d topological
Belavin--Polyakov soliton in isotropic ferromagnet. Analytical
solutions of the scattering problem are constructed: (i) exactly
for any magnon wave vectors for the partial wave with the
azimuthal number $m=1$ (translational mode), and (ii) in the
long-- and short--wave limits for the rest modes. The magnon mode
frequencies are found for the finite size magnets. An effective
equation of the soliton motion is constructed. The magnon density
of states, connected with the soliton--magnon interaction, is
found in a long--wave approximation.
\end{abstract}
\pacs{75.10.Hk, 75.30.Ds, 75.50.Ee} %
}

Nonlinear topologically nontrivial excitations (solitons) are
well--known to play a special role in a low--dimensional magnetic
systems. For example, the presence of vortices in 2d easy--plane
(EP) magnets gives rise to Berezinski\u\i--Kosterlitz--Thouless
phase transition \cite{Berezinsky72&Kosterlitz73}. Kinks in 1d
magnets and localized Belavin--Polyakov solitons (BP--solitons
\cite{Belavin75}) in 2d isotropic magnets are responsible for the
destruction of the long--range order at finite temperature. This
can be explained within the scope of so--called soliton
phenomenology, where the magnet can be described as a
two--component gas of elementary excitations: solitons and
magnons. Such approach was developed for 1d magnets
\cite{Currie80}, see also Refs.
\cite{Mikeska91&Wiesler94,Ivanov95e}. The soliton signature in
dynamical response functions can be observed experimentally.
Translational motion of solitons leads to the so--called soliton
central peak. Another possibility to detect the soliton signature
is to look for magnon modes, localized on the soliton ({\em local
modes}, LM), observed in \cite{Boucher87}. For 1d magnets such
scattering causes the change of the magnon density of states which
is necessary for self--consistent calculation of temperature
dependence of the soliton density \cite{Currie80}.

For 2d the concept of soliton--magnon gas has been extended, e.g.
to describe EP magnets \cite{Mertens87&Mertens89}, and to explain
the EPR line--width in easy--axial magnets
\cite{Waldner83&Waldner86&Waldner92,Zaspel93&Zaspel95&Zaspel96&Subbaraman98}.
However the general behaviour of the 2d soliton dynamics is not
clear at present. In particular, the form of inertial terms in the
dynamical equations for the soliton centre is unknown; the soliton
density have not been calculated, but been used as input
parameter.

The problem of the soliton dynamics or the problem of LM existing
are intimately connected with the soliton--magnon scattering. For
example, using numerical data for the scattering amplitude the
non--Newtonian effective equation of motion of the magnetic vortex
was constructed \cite{Ivanov98}. For the 2d EP antiferromagnet
(AFM), finite--frequency truly localized internal mode was
predicted \cite{Ivanov96}. For all mentioned papers, an analysis
of 2d solitons was carried out numerically. It becomes especially
important to analyze models for which analytical results can be
obtained. We are aware of only the exact solution for BP--soliton
\cite{Belavin75}, which describes a topological soliton in
isotropic magnets.

In this Letter we have investigated the soliton--magnon scattering
for the BP--soliton in the isotropic ferromagnet (FM). An exact
analytical solution of the scattering problem is constructed. An
effective equation of the soliton motion is found for the finite
size FM. We also analyzed the long-wave asymptotics of the magnon
density of states in the presence of BP--soliton.

\paragraph*{Solitons and Magnons.}

The dynamics of the FM follows the Landau--Lifshitz equations for
the normalized magnetization $\vec m$ \cite{Kosevich90}. In
angular variables, $ m_x+im_y=\sin \theta \exp (i\phi )$, these
equations correspond to the Lagrangian $L=(A/2)\int d^2x{\cal L}$,
\begin{equation} \label{eq:L}
{\cal L}=\frac 2D\left(1-\cos\theta\right) \frac{\partial \phi
}{\partial t}-\left( \nabla \theta \right) ^2-\left( \nabla \phi
\right) ^2\sin ^2\theta
\end{equation}
where the $A=JS^2$, $J$ is the exchange integral, and $S$ the
atomic spin. The magnon solutions have a form $\theta
=\text{const}\ll 1$, $\phi =\vec{k}\cdot\vec{r}-\omega(k)t$ with
the dispersion law $\omega(k)=Dk^2$, so the constant $D$ is the
spin--wave stiffness.

The simplest static nonlinear excitations in 2d case are the
BP--solitons \cite{Belavin75}:
\begin{equation}
\tan \frac{\theta _0}2=x^{-|\nu |},\qquad \phi _0=\varphi _0+\nu
\chi, \qquad x=\frac rR.  \label{eq:BP}
\end{equation}
Here $r$, $\chi$ are the polar coordinates in the magnet plane,
the integer $\nu$ determines $\pi_2$ topological charge of the
soliton, and $R$ , $\varphi_0$ are arbitrary parameters. The
energy of the soliton (\ref{eq:BP}) $E_0=4\pi A|\nu|$ does not
depend on the soliton's radius $R_0$ due to the conformal
invariance of the static model \cite{Belavin75}. Apparently, such
invariance breaks for FM in the dynamical case.

\paragraph*{The scattering problem.}

The equations for linear oscillation on the soliton background,
$\theta =\theta _0(r)+\vartheta (r,\chi,t)$ and $\phi=\nu\chi+
\varphi_0+(\sin\theta_0)^{-1}\mu(r,\chi,t)$ can be represented in
the form of the single equation for the complex quantity $\psi
=\vartheta+i\mu$,
\begin{equation} \label{eq:psi}
\left(-\nabla^2+\frac{\nu ^2}{r^2}\cos 2\theta _0\right)
\psi-\frac{ 2i\nu }{r^2}\cos \theta _0\frac{\partial\psi
}{\partial\chi}=\frac iD \frac{\partial\psi}{\partial t}.
\end{equation}
The solution of Eq. (\ref{eq:psi}) has the form of a superposition
of cylindrical waves $\psi =\sum_{m=-\infty }^\infty f_m e^{im\chi
-i\omega t}$, $m$ is the azimuthal quantum number. The partial
waves $f_m$ satisfy the 2d--radial Schr\"odinger eigenvalue
problem (EVP),
\begin{mathletters} \label{eq:EVP4f}
\begin{eqnarray} \label{eq:f_m}
&& \hat{H}f_m = (kR)^2f_m, \quad \hat H=-\nabla _x^2+U_m(x) \\
\label{eq:U_m} && U_m(x) =(1/x^2) \left(m^2+2m\nu \cos \theta _0+\nu
^2\cos 2\theta _0\right).
\end{eqnarray} \end{mathletters}
Far from the soliton, $r\gg R$, the function $U_m\approx
p^2/x^2-4\nu(m+2\nu)(1/x)^{2(\nu+1)}$, so
\begin{equation} \label{eq:f_m_scat}
f_m\propto J_{|p|}(z)+\sigma _m^\nu
Y_{|p|}(z)+4\nu(m+2\nu)(kR)^{2\nu}F_{|p|}(z),
\end{equation}
where $z=kr$, $J_{|p|}$ and $Y_{|p|}$ are Bessel and Neumann
functions, respectively, $p=\nu+m$. The quantity $\sigma _m^\nu =
-\tan \delta_m^\nu $ determines the amplitude of the
soliton--magnon scattering, the $S$-matrix can be written as
$S_m^\nu=\exp(2i\delta _m^\nu)$. The correction $F$, which is
small at $kR\ll1$, has a power decay $F\propto z^{-5/2}$ at
$kr\gg1$, but its contribution is important to calculate the
scattering amplitude. The explicit form of $F$ can be found after
long but simple exercises with the unhomogeneous Bessel equation.

In the case of $\omega=0$ there is an exact solution \cite{Ivanov95g} of EVP
(\ref{eq:EVP4f}) due to the restoration of the conformal invariance
of the model :
\begin{equation} \label{eq:f_m^0}
f_m^{(0)}=x^{-m}\sin \theta _0
\end{equation}
(we shall discuss  the case $\nu>0$; for analyzing of $\nu <0$, it
is sufficient to replace $m$ by $-m$). It should be noted that
solutions (\ref{eq:f_m^0}) are finite and have no singularities at
$r\to0$, as well as at $r\rightarrow \infty $ when $-|\nu |<m\leq
|\nu |$. This at once leads to the existence of $2|\nu |$ LM with
zero frequencies (zero LMs). The physical meaning of two such
modes is obvious: the translational mode $f_{m=1}^{(0)}$ describes
the displacements of a soliton as a whole; the mode
$f_{m=0}^{(0)}$ corresponds to the change of $\varphi _0$ and
soliton radius $R$. Note that LMs are limit points for magnon
modes of the continuous spectrum when $k\to 0 $ (unlike 1d magnet,
in which LMs are separated from the continuous spectrum by the gap
\cite{Ivanov95e}).

For $m>\nu $, the solution $f_m^{(0)}$ has a singularity in the
origin, but in this case the second independent solution of EVP
(\ref{eq:EVP4f}) with $k=0$ , regular at $r\to0$, $f_m^{(1)}
=x^m\sin\theta_0\left[x^{2\nu}/(m+\nu)+2/m+x^{-2\nu}/(m-\nu)
\right],$ can be used. Thus, for $\omega =0$, we can construct at
least one solution $f_m^{(0)}$ or $f_m^{(1)}$ of the EVP
(\ref{eq:EVP4f}) that does not have a singularity in the origin.

The existence of the exact solution $f_m^{(0)}$ permits to present
the Schr\"odinger operator $\hat H$ in the factorized form $\hat
H=\hat A^{\dag } \hat A$, $\hat Af_m^{(0)}=0,$ where
\begin{equation} \label{A&A^dag}
\hat A=-\frac d{dx}+\frac{f{_m^{(0)}}^{\prime }}{f_m^{(0)}},\text{
}\hat A ^{\dag }=\frac d{dx}+\frac 1x+\frac{f{_m^{(0)}}^{\prime
}}{f_m^{(0)}}.
\end{equation}
Introducing the function $g_m=\hat Af_m$, the EVP (\ref{eq:EVP4f})
can be rewritten as
\begin{mathletters} \label{eq:EVP4g}
\begin{eqnarray} \label{eq:g_m}
&& \hat {{\cal H}}g_m=(kR)^2g_m, \quad \hat{{\cal H}}\equiv \hat
A\hat A ^{\dag }=-\nabla _x^2+{\cal U}_m \\ \label{eq:U4g} &&
{\cal U}_m(x)=\frac{n^2}{x^2}-\frac{4\nu (m-1)}{x^2}
\sin^2\frac{\theta _0}2,\; n=\nu +m-1.
\end{eqnarray} \end{mathletters}
The function $f_m$ can be found via $g_m$ as $f_m=(kR)^{-2}\hat
A^{\dag }g_m$ . Using such transformation one can find an exact
solution for the translational mode. Indeed, in this case ${\cal
U}_1(x)=\nu ^2/x^2$, so $ g_1=-J_\nu (kr)$ and
\begin{equation} \label{eq:f_1}
f_1=J_{\nu +1}(kr)-\frac{2\nu }{kr}\cdot \frac{J_\nu
(kr)}{(r/R)^{2\nu }+1}.
\end{equation}
The existence of the exact solution for any $k$ is the unique
property of the model (\ref{eq:L}). It demonstrates the fact,
mentioned above (\ref{eq:f_m_scat}), that the deviation $f_1$ from
the asymptote $J_{\nu +1}$ is specified by  power decaying apart
from the soliton (while the deviations far from the vortex core
are decaying exponentially \cite{Ivanov96,Ivanov98}). Again, the
EVP (\ref{eq:EVP4g}) is more convenient for numerical  analysis
because the ``potential'' ${\cal U}_m$ is repulsive.

In order to describe the magnon scattering in the long--wave
approximation we use the fact that exact zero functions
$f_m^{(0)}$ or $f_m^{(1)}$ describe correctly the solution in the
region $r\ll1/k$. For small but finite values of $k$ we make the
ansatz
\begin{equation} \label{eq:g-via-f}
\hat A^{\dag }g_m=(kR)^2f_m^0,
\end{equation}
where $f_m^0$ is one of the exact zero solutions, seen above. On
the other hand we can use the scattering approximation with
account of the correction $F$ (\ref{eq:f_m_scat}). Therefore, over
a wide range of values of $r$, for $R\ll r\ll 1/k$, we can use the
ansatz (\ref{eq:g-via-f}): on one hand, starting from the exact
zero solution, and on the other hand explaining the results in
term of the scattering problem.  Awkward calculations lead to
long--wave asymptotics of the scattering amplitude $\sigma _m^{\nu
=1}$ for the soliton with $\nu=1$ having minimal energy. It was
found that the scattering intensity is maximal for the LM with
$m=0$,
\begin{mathletters} \label{eq:sigma}
\begin{equation} \label{eq:sigma4m=0}
\sigma _0(k)=\frac \pi {2\ln \left( 1/kR\right) }, \quad kR\ll1.
\end{equation}
The same dependence occurs for solitons with $\nu\neq1$ at
$m=-\nu+1$.

The second LM (translational mode, $m=1$) does not
scatter in accordance with Eq. (\ref{eq:f_1}). For the quasi--LM
$(m=-1)$ we can write $\sigma _{-1}(k)=\pi (kR)^2\ln \left( 1/kR\right)
$. In the range of absence of LM, we can restore the general dependence
\begin{equation} \label{eq:sigma4m}
\sigma _m(k)=-\frac{\pi (kR)^2}{2|m|(m+1)}, \quad kR\ll1.
\end{equation}

In the short--wave limit $ k\gg |m|/R$  we can use WKB approximation for
all $r$ far from the turnover point, $|m|/k\ll r$, and at the
same time  the asymptotic
$ g_m\propto J_{|\nu -m+1|}(cr)$ at $r\ll R$. Thus at a wide range of
values of $r$, $|m|/k\ll r\ll R$, both solutions are valid. Their
comparison gives:
\begin{equation} \label{eq:sigma(k>>1)}
\sigma _m^\nu (k)= \frac{\pi (m-1)}{\sin \left( \pi /2\nu \right)
} \cdot \frac 1{kR},\quad kR\gg |m|.
\end{equation} \end{mathletters}
One can see that $\sigma_m$ tends to zero for both long-- and
short--wave limits, but with opposite signs. Due to this, the
values of $\delta _m^\nu $ at $k\to\infty$ and $k\to0$ differ by
$\pm\pi$, the sings ``$+$'' and ``$-$'' correspond to $m>1$ and
$m<0$, respectively. Therefore, the scattering amplitude has a
pole at some finite $k=k_p$, the value of $k_p$ is growing as
$|m|$ with the growth of $|m|$, which is confirmed by the
numerical calculations. Note that for the magnon scattering by 1d
soliton in a number of models the values of $\delta$ at
$k\to\infty $ and $k\to0$ differ by $-\pi$, that causes the
decrease of the total number of magnon states by one
\cite{Ivanov95e}.

\paragraph*{Finite size magnet.}

The investigation of the magnon eigenmodes on the soliton in the
finite size magnet can be used for the analytical description of
the direct numerical simulation data of the soliton motion. In
this way the non--Newtonian vortex dynamics was explained in
\cite{Ivanov98} for EP FMs. On the other hand such calculation can
be used to describe eigenmodes for small FM particles in
so--called vortex state \cite{Usov93}.

We consider magnon modes in the circular system of radius $L$ with
the soliton in the centre. Different boundary conditions (BC),
both fixed (Dirichlet BC, DBC, $f_m\bigr|_{r=L}=0$) and free
(Neumann BC, NBC $\left( \partial f_m/\partial r\right)
\bigr|_{r=L}=0$) will be discussed. Without soliton the magnon
spectrum in such a magnet is discrete, e.~g. for DBC the
eigenvalues $k_{m,i}=j_{m,i}/L$, where $j_{m,i}$ is the $i-$th
zero of $ J_m$. If the soliton is present, this is true again for
some modes with small scattering $\sigma _m\ll 1$
\cite{Ivanov96,Ivanov98}, $ k_{m,i}=z_{m,i}/L$, where $z_{m,i}$ is
the $i-$th root of the equation $ J_{|p|}(z)+\sigma Y_{|p|}(z)=0$.

However, in the LM case $-\nu<m\leq\nu $, the system has higher
symmetry due to the restoration of the conformal invariance.
Naturally, in the finite size magnet case there are
quasi--Goldstone modes (qGMs) with anomalously small frequencies,
i.~e. $kL\ll 1$. In particular, the qGM appears for the
translational motion ($|m|=1$) of the vortex in the EP FM. The
existence of such a mode is determined by the scattering only
\cite{Ivanov96,Ivanov98}.

We discuss the most important case of qGM with $m=1$, which
describes the translational motion of the soliton and could be
used for the construction of the dynamical equations of the
soliton \cite{Wysin96,Ivanov98}. Using the exact solution
(\ref{eq:f_1}) in the region $kR\ll 1$ one can find the frequency
of the translational qGM in the form
\begin{equation}\label{eq:omega_0}
\omega_0=\pm8DR^2L^{-4}
\end{equation}
signs ``$-$'' and ``$+$'' correspond to DBC and NBC, respectively.
The next translational mode has a frequency $\omega _1=Dj^2L^{-2}$
or $\omega _1=-Dj^{\prime }{}^2L^{-2},$ respectively. The highest
modes are separated from these ones by the finite gap, so the
picture of doublets, which is characteristic for the
vortex--magnon scattering, is absent \cite{Ivanov98}. Thus it is a
belief that the account of these two modes gives the adequate
description of soliton motion.

To describe the soliton dynamics with two characteristic frequencies
we can use the 2$^{nd}$ order equation
\begin{equation} \label{eq:FM}
M\frac{\partial ^2\vec X}{\partial t^2}+G\left[ \vec e_z\times
\frac{\partial \vec X}{\partial t}\right] =\vec F,\quad \vec{F}
=\mp\frac{32\pi AR^2\vec{X}}{L^4},
\end{equation}
where $M$ and $G$ are effective mass and the constant of
gyroforce, $G=4\pi\nu A/D$ is determined by the topology only
\cite{Thiele73&Nikiforov83&Huber82,Ivanov95e}, $\vec{F}$ is an
image--force acting on soliton due to the boundary. Such a form of
$\vec{F}$ can be explained by taken into account that the magnetic
vortices interact as 2d charges, and the BP--soliton with $\nu=1$
can be interpreted as a vortex dipole.

Assuming $\omega _0\ll \omega _1$, it is easy to compare $\omega
_0$, $ \omega _1$ with the frequencies of the Eq. (\ref{eq:FM})
and extract the unknown value of the effective mass (this method
was used for the vortices in EP magnets with the using of the
numerical values of $\omega _0,\omega _1$ as an input
\cite{Wysin96}). Indeed, $\omega _0\approx -\alpha /G$ coincides
with the frequency of qGM. Comparing $\omega _1$ with the value $
\omega_1\approx-G/M$ we obtain
\begin{equation}  \label{eq:M}
M_D=-\frac{4\pi A}{D^2}\cdot \left( \frac Lj\right) ^2,\quad
M_N=\frac{4\pi A}{ D^2}\cdot \left( \frac L{j^{\prime }}\right)^2
\end{equation}
for DBC and NBC, respectively. Therefore $M$ is unlocal as the
coefficient $G_3$ in $3^{rd}$-order equations of motion for the FM
\cite {Ivanov98}. Note that the dependence $M\propto L^2$ agree
with the result $ M\propto 1/K$, $K$ is the anisotropy constant,
for the easy--axis FM \cite{Ivanov89}, if the characteristical
length $\Delta _0=\sqrt{A/K}$ is changed by $L$.

The analysis shows that qGMs appear for all higher $\nu$ in the
region of the existence of LMs, $-\nu<m\leq\nu$. Their frequencies
are proprotional to $(D/L^2)(R/L)^{2|\nu+m-1|}$ or $\left[D/L^2\ln
(L/R)\right]$.

\paragraph*{Magnon density of states.}

The main point of the 1d soliton (kink) phenomenology
\cite{Currie80} is a change in the magnon density of states due to
the presence of the kink, $\rho _{1d}(k)=\left( 1/2\pi \right)
d\delta(k)/dk$, where $\delta (k)$ is a phase shift by
kink--magnon interaction. It causes the decrease of the total
number of magnon states, so the magnons free energy changes
accordingly.

Let us transfer this approach to the 2d case. Using the magnon
density of states the soliton phenomenology of 2d magnets could be
constructed. In particular, the soliton density could be
calculated. The free 2d magnon has an expansion in terms of
cylindrical waves, $J_m(kr)e^{im\chi }$, with the quantized
angular part. Hence only the radial part $J(kr)$ must be
quantized. In the circular geometry with the radius $L$ the
eigenvalues $k_n$ satisfy the conditions $k_nL= j_{m,n}$ for the
DBC. For $n\gg1$ the standard condition $k_nL \approx\pi n$ takes
place. However, for $|m|\gg 1$ the $1^{st}$ zero
$j_{m,1}\approx|m|$, so the set of allowed values of $m$ must be
limited by the condition $|m|<kL$. Accordingly, the 2d density of
states is
\begin{equation} \label{eq:rho(k)}
\rho(k)=\frac 1\pi \sum_{m=-kL}^{kL}\frac{d\delta _m(k)}{dk}.
\end{equation}
For small $k$, the function $\rho (k)$ diverges due to the mode with
$m=0$. This contribution is dominant for low temperatures. Using
(\ref{eq:sigma4m=0}) one can obtain $\rho (k)\approx -\left(
2k\right)^{-1}\ln^{-2}\left(kR\right)$. However, our numerical
calculations done for largest $k$ shows that the series
(\ref{eq:rho(k)}) is alternating and the total number of magnon states
does not decrease as for 1d. There is only a redistribution of the
magnon modes between the states with the opposite $m$.

\paragraph*{Conclusion.}

We have constructed an exact analytical solution of the
soliton--magnon scattering problem for the isotropic FM.  We used
these results for different properties of solitons and LMs. In
particular, LM frequencies have been calculated for the finite
size FM. Eigenmodes with anomalous small frequencies appear in the
small FM particle having cylindrical geometry with the soliton in
the centre. The effective equation of the soliton motion have been
constructed. It was analyzed the magnon density of states in the
presence of the soliton.

All the results for the magnon modes (obtained for the FM) can be
expanded to the Lorentz--invariant nonlinear $\sigma$--model.
Lagrangians for this model results from (1) after replacing of the
gyroscopical term with $1/D$ by ($1/c^2)(\partial \vec l/\partial
t)^2$, where $c$ is the characteristic speed. For magnets, this
model can be used to describe the AFM dynamics, where the
dispersion law $\omega_{FM}(k)=Dk^2$ must be replaced by
$\omega_{AFM}=\pm ck$. Two lowest modes with $m=1$ having the
frequencies $\omega_0=\pm2\sqrt2cRL^{-2}$ describe two independent
modes of translational dynamics of AFM BP--soliton with $G=0$, the
effective mass $M_{AFM}=E_0/c^2$ under the action of image soliton
force. The magnon density of states $\rho(k)$ has the same form as
for the FM.

We may use these results for the Euclidean version of the
$\sigma$--model, important for the quantum chains with the AFM
interaction. Properties of such chains are connected with the {\em
instantons} of the $\sigma$--model. Both instantons with the
BP--soliton structure \cite{Fradkin} and {\em merons} with
half--integer topological charge \cite{Affleck} are discussed. The
knowledge of the total set of eigenvalues on the instanton
background, especially zero modes, is necessary to calculate the
fluctuation determinant.

The authors are indebted to V.~G.~Bar'yakhtar, G.~Holzwarth,
D.~I.~Sheka for stimulated discussions. The research was supported
in part by a Grant No 2.4/27 from the Ukrainian State Foundation
for Fundamental Research. D.~Sh. acknowledges the support of ISSEP
Grant No YSU082065.


\begin{references}
\bibitem[\S]{Sheka:e-mail}  e-mail: sheka@rpd.univ.kiev.ua

\bibitem{Berezinsky72&Kosterlitz73}  V.~L. Berezinski\u\i, Sov. Phys JETP
{\bf 34}, 610 (1972); J.~M. Kosterlitz and D.~J. Thouless, J.
Phys. {\bf C 6}, 1181 (1973).

\bibitem{Belavin75}  A.~A. Belavin and A.~M. Polyakov, JETP Lett. {\bf 22},
245 (1975).

\bibitem{Currie80}  J.~F. Currie, J.~A. Krumhansl, A.~R. Bishop, and S.~E.
Trullinger, Phys. Rev. {\bf B 22}, 477 (1980).

\bibitem{Mikeska91&Wiesler94}  H.~J. Mikeska and M. Steiner, Adv. Phys. {\bf
40}, 191 (1991).

\bibitem{Ivanov95e} V. G. Bar'yakhtar and B. A. Ivanov, Sov. Sci.
Rev. Sec.A. --- Phys. Reviews  ed. by I. Khalatnikov,{\bf 16}, No.
3, Amsterdam (1993); B.~A. Ivanov and A.~K. Kolezhuk, Low Temp.
Phys. {\bf 21}, 275 (1995).

\bibitem{Boucher87}  J.~P. Boucher, G. Rius, and Y. Henry, Europhys. Lett.
{\bf 4}, 1073 (1987).

\bibitem{Mertens87&Mertens89}
F.~G. Mertens, A.~R. Bishop, G.~M. Wysin, and C. Kawabata, Phys.
Rev. Lett.  {\bf 59},  117  (1987); Phys. Rev. {\bf B39},
591  (1989).

\bibitem{Waldner83&Waldner86&Waldner92}
F. Waldner, J.~Magn. Magn. Mater. {\bf 31---34},  1203  (1983);
{\bf 54---57},  873  (1986); {\bf 104---107},  793  (1992).

\bibitem{Zaspel93&Zaspel95&Zaspel96&Subbaraman98}
C.~E. Zaspel, T.~E. Grigereit, and J.~E. Drumheller, Phys. Rev. Lett.
{\bf 74}, 4539  (1995); C.~E. Zaspel and J.~E. Drumheller, Int. J. Mod.
Phys. {\bf 10},  3649  (1996).

\bibitem{Ivanov98}  B.~A. Ivanov, H.~J. Schnitzer, F.~G. Mertens, and G.~M.
Wysin, Phys. Rev. {\bf \ B 58}, 8464 (1998).

\bibitem{Ivanov96}  B.~A. Ivanov, A.~K. Kolezhuk, and G.~M. Wysin, Phys.
Rev. Lett. {\bf 76}, 511 (1996).

\bibitem{Kosevich90}  A.~M. Kosevich, B.~A. Ivanov, and A.~S. Kovalev, Phys.
Rep. {\bf 194}, 117 (1990).

\bibitem{Ivanov95g}  B.~A. Ivanov, JETP Lett. {\bf 61}, 917 (1995).

\bibitem{Usov93}  N.~A. Usov and S.~E. Peschany, J.~Magn. Magn. Mater. {\bf
118}, L290 (1995).

\bibitem{Wysin96} G.~M. Wysin, Phys. Rev. {\bf B 54}, 15156 (1996).

\bibitem{Thiele73&Nikiforov83&Huber82}  A.~A. Thiele, Phys. Rev. Lett.
{\bf 30}, 239 (1973); A.~V. Nikiforov and E.~B. Sonin, Sov. Phys JETP
{\bf 58}, 373 (1983); D.~L. Huber, Phys. Rev. {\bf B 26}, 3758 (1982).

\bibitem{Ivanov89}  B.~A. Ivanov and V.~A. Stephanovich, Phys. Lett.
{\bf A 141}, 89 (1989).

\bibitem{Fradkin} E.~Fradkin, {\em Field theories of condensed
matter systems}, in {\em Frontiers in Physics}, {\bf 82},
Addison--Wesley (1991).

\bibitem{Affleck} Ian Affleck, J. Phys.: Condens. Matter. {\bf 1}, 3047
(1989).

\end{references}
\end{document}